\documentclass[%
 reprint,
 amsmath,amssymb,
 aps,
 pra,
]{revtex4-2}

\usepackage[qm]{qcircuit}
\usepackage{graphicx}
\usepackage{dcolumn}
\usepackage{bm}
\usepackage{hyperref}
\usepackage[mathlines]{lineno}

\begin{document}


\title{Procedure for improving cross-resonance noise resistance using pulse-level control}

\author{David Danin$^{1}$, Felix Tennie$^{1,2}$}
\affiliation{$^{1}$University of Oxford, Department of Physics, Clarendon Laboratory, Parks Road, Oxford OX1 3PU, United Kingdom of Great Britain and Northern Ireland \\
$^{2}$Imperial College London, Department of Aeronautics, South Kensington Campus,
London, SW7 2AZ, United Kingdom of Great Britain and Northern Ireland}


\date{\today}

\begin{abstract}

Current implementations of superconducting qubits are often limited by the low fidelities of multi-qubit gates. We present a reproducible and runtime-efficient pulse-level approach for calibrating an improved cross-resonance gate CR($\theta$) for arbitrary $\theta$. This CR($\theta$) gate can be used to produce a wide range of other two-qubit gates via the application of standard single-qubit gates. By performing an interleaved randomised benchmarking experiment, we demonstrate that our approach leads to a significantly higher noise resistance than the circuit-level approach currently used by IBM. Hence, our procedure provides a genuine improvement for applications where noise remains a limiting factor.

\end{abstract}

\maketitle



Quantum computers promise to provide unprecedented computational power in applications such as optimisation or simulation by exploiting the fact that information is not encoded in classical but in quantum systems \cite{Montanaro_2016, Nielsen_Chuang_2010}. In recent years, the field has seen the rapid development of improved quantum hardware \cite{Brooks_2023}. However, the practical benefit of commercially available quantum computers based on superconducting qubits remains limited by the relatively low fidelities of multi-qubit interactions \cite{Kjaergaard_2020}.

In addition to the improvement of hardware components, it is possible to enhance gate fidelities using optimised control approaches \cite{Glaser_2015}. With the introduction of Qiskit Pulse \cite{Qiskit}, it is now possible to precisely control real quantum hardware via the IBM Quantum Lab \cite{IBMQuantumLab}. As explained in Ref.~\cite{Alexander_2020}, one can specify the amplitude, frequency and phase of the physical microwave pulses that drive the qubits to implement custom single-qubit and multi-qubit gates \cite{Krantz_2019}. Hence, Qiskit Pulse allows for designing and testing control approaches on the level of physical operations instead of logical operations \cite{Gambetta_2017}.

\begin{figure}
    \centering
    \includegraphics[width=\linewidth]{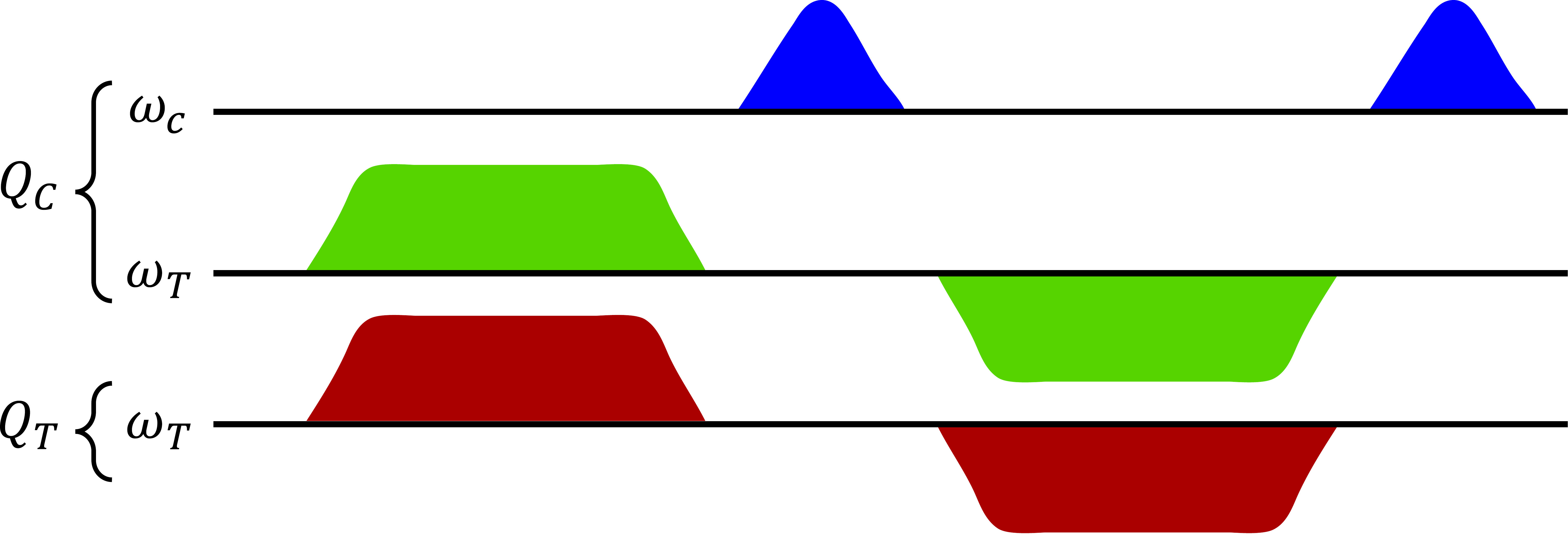}
    \caption{The schematic adapted from Ref.~\cite{Sheldon_2016} illustrates the pulse schedule for a CR($\pi/2$) cross-resonance gate. The control qubit $Q_C$ is driven at the resonant frequency $\omega_T$ of the target qubit $Q_T$. Unwanted terms in the interaction Hamiltonian $H_I$ are suppressed by the echo sequence on $Q_C$ (i.e. the upper two drive lines) and cancellation tones on $Q_T$ (i.e. the lower drive line), such that $H_I \approx g(Z \otimes X)$ follows.}
    \label{fig:pulse_schematic}
\end{figure}

The cross-resonance gate (CR) is a particularly important two-qubit interaction on superconducting qubits, as it combines various desirable features and enables the construction of the Controlled-NOT gate \cite{Rigetti_2010} which is the standard entangling operation of universal gate sets \cite{Nielsen_Chuang_2010}. As demonstrated in Ref.~\cite{Sheldon_2016}, one can implement a high-fidelity CR gate using the pulse sequence schematically illustrated in Fig.~\ref{fig:pulse_schematic}. When successfully calibrated, this corresponds to implementing the interaction $H_I \approx g (Z \otimes X)$ with $g$ some coupling constant that depends on the hardware components and the drive amplitude \cite{Sheldon_2016, Magesan_2020}. The time evolution operator generated by this Hamiltonian reads $U(t) = \cos(gt) \mathbf{1} \otimes \mathbf{1} - i \sin(gt) Z \otimes X$ \cite{Krantz_2019}. Setting $gt = \pi/4$ by changing the amplitude or duration of the pulses, a CR($\pi/2$) gate is implemented \cite{Sheldon_2016}.

By using Qiskit Pulse on publicly available quantum backends via the IBM Quantum Lab, the method presented in Ref.~\cite{Sheldon_2016} can be extended to significantly improve the noise resistance of multi-qubit gates. Specifically, we describe a pulse-level approach for calibrating a set of cross-resonance gates and demonstrate that they achieve significantly higher noise resistances than their circuit-level implementations used by IBM. Crucially, the procedure we present is straightforwardly replicated. Accordingly, we provide a powerful extension to the set of high-fidelity, multi-qubit gates on currently available quantum computers based on superconducting qubits.

First, we introduce a runtime-efficient procedure for calibrating a $Z \otimes X$ cross-resonance interaction CR($\theta$) via the IBM Quantum Lab, thereby extending the approach presented in Ref.~\cite{Sheldon_2016} to values of $\theta$ other than $\pi/2$. Second, we describe how this CR($\theta$) gate can be used to straightforwardly implement a range of other two-qubit interactions. And finally, we demonstrate that our pulse-level implementation achieves significantly higher noise resistances, compared to the circuit-level implementation which IBM currently uses, by performing a modified interleaved randomised benchmarking experiment \cite{Magesan_2012}.

We begin by presenting our procedure for calibrating a CR($\theta$) gate. The method is adapted from Ref.~\cite{Sheldon_2016} but differs in two important respects. First, we generalise the procedure to values of $\theta$ other than $\pi/2$. And second, we streamline the procedure to make it more runtime-efficient. This enables us to perform the full calibration procedure on publicly available quantum backends via the IBM Quantum Lab, even under runtime constraints.

\begin{figure}
    \centering
    \includegraphics[width=\linewidth]{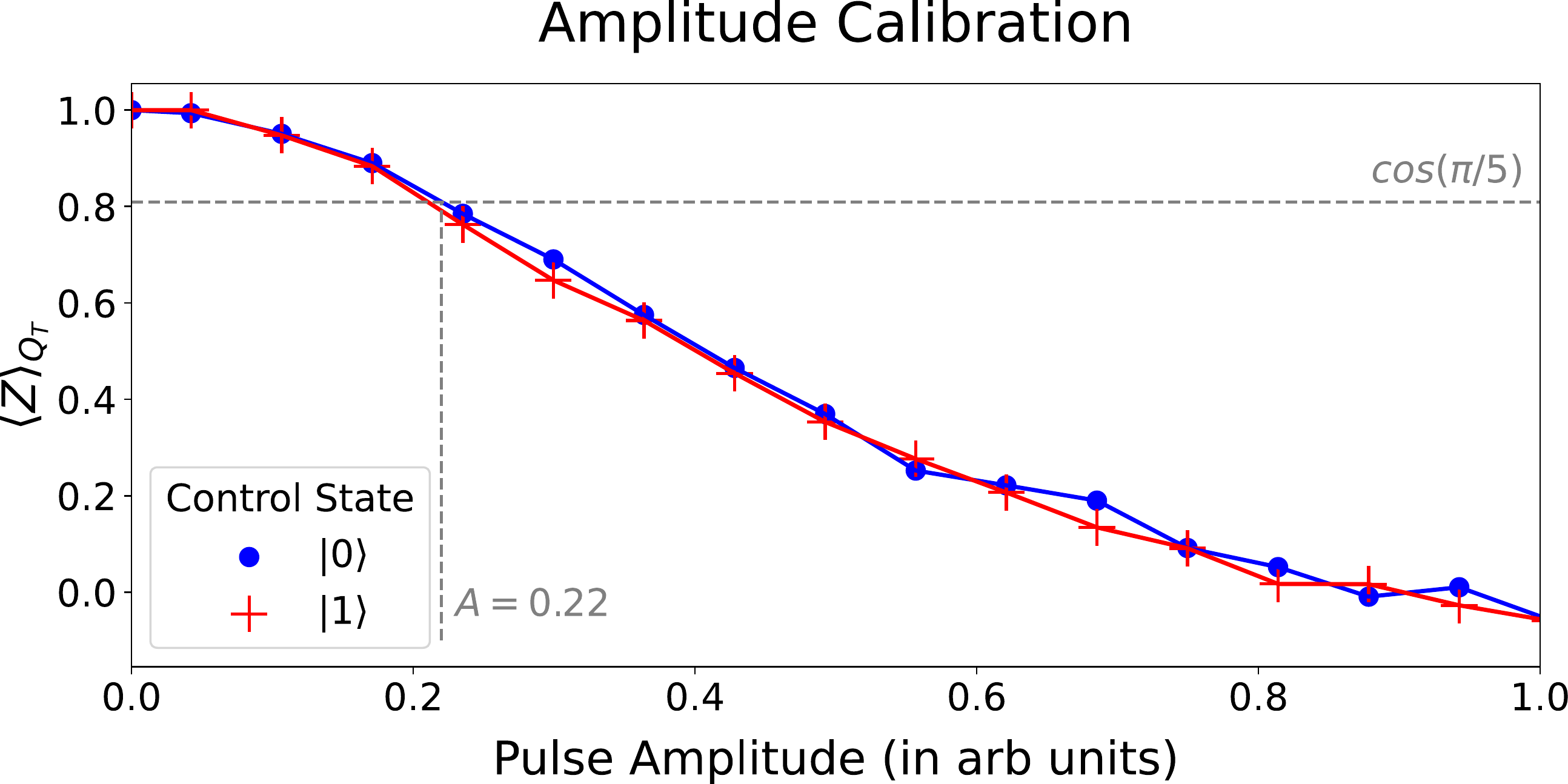}
    \caption{Results for the amplitude calibration experiment with $\theta = \pi/5$. We determine the correct pulse amplitude by reading off $A_\theta$ which is defined by $\langle Z(A_\theta) \rangle = cos(\theta)$. Using this amplitude will lead to a CR($\pi/5$) gate which has the same duration as the CR($\pi/2$) pulse that forms part of the standard Controlled-NOT implementation between $Q_C$ and $Q_T$. The error bars are smaller than the marker size.}
    \label{fig:amplitude_calibration}
\end{figure}

First, we need to determine the correct amplitude for the CR($\theta$) pulse between our control qubit $Q_C$ and target qubit $Q_T$. For this, we define a flat-top pulse with Gaussian edges and some real amplitude $A$. The width and Gaussian rise time of the pulse are inherited from the CR($\pi/2$) pulse that forms part of the standard Controlled-NOT implementation between $Q_C$ and $Q_T$. While testing these parameters might lead to a more precise calibration, we adopt this assumption to significantly reduce the calibration runtime. We note that this assumption is self-consistent since it leads to a high-fidelity CR($\theta$) gate as shown in the subsequent experiments.

Then, we sweep through different real amplitude values $A$ and measure $Q_T$ in the computational basis to calculate the Pauli expectation value $\langle Z(A) \rangle$. We repeat the experiment with $Q_C$ initialised in $|0\rangle$ and $|1\rangle$. Assuming that the $Z \otimes X$ or $Z \otimes Y$ component in $H_I$ is much larger than the other contributions, we find $\langle Z \rangle \approx \cos(\theta)$, as for an ideal CR($\theta$) gate we have $\langle Z \rangle = \cos(\theta)$. Note that the assumption made here is consistent with the results of the subsequent tomography experiments. Hence, for a given $\theta$, we can find the amplitude $A_\theta$ that leads to the correct value of $\langle Z \rangle$ and use this amplitude for our pulse.

Second, we need to determine the correct phase for the CR($\theta$) pulse. For this, we sweep through different pulse widths with our flat-top Gaussian pulse using the real amplitude that we previously determined. We repeat the experiment with $Q_C$ initialised in $|0\rangle$ and $|1\rangle$. By measuring the expectation values $\langle X \rangle$, $\langle Y \rangle$, and $\langle Z \rangle$ on the target qubit, we reconstruct the coefficients of the terms in the cross-resonance interaction Hamiltonian $H_I$. For details regarding the Hamiltonian tomography experiment, we refer the reader to Ref.~\cite{Sheldon_2016} and Ref.~\cite{Hamiltonian_Tomography_2022}.

Hence, we can determine the coefficients $C_{ZX}$ and $C_{ZY}$ of the cross-resonance $Z \otimes X$ and $Z \otimes Y$ components in $H_I$, respectively. Recognising that $C_{ZX} \propto \cos(\phi-\phi_0)$ and $C_{ZY} \propto \sin(\phi-\phi_0)$, where $\phi$ is the phase of the cross-resonance pulse \cite{Rigetti_2010}, we can set the phase of the pulse to $\phi_0 = - \tan^{-1}(C_{ZY} / C_{ZX})$ such that the $Z \otimes Y$ component in $H_I$ vanishes. Thereby, we can calibrate the phase of the cross-resonance pulse in a single experiment. This provides a far more efficient method than sweeping through phases as described in Ref.~\cite{Alexander_2020} and Ref.~\cite{Sheldon_2016}.

\begin{figure}
    \centering
    \includegraphics[width=\linewidth]{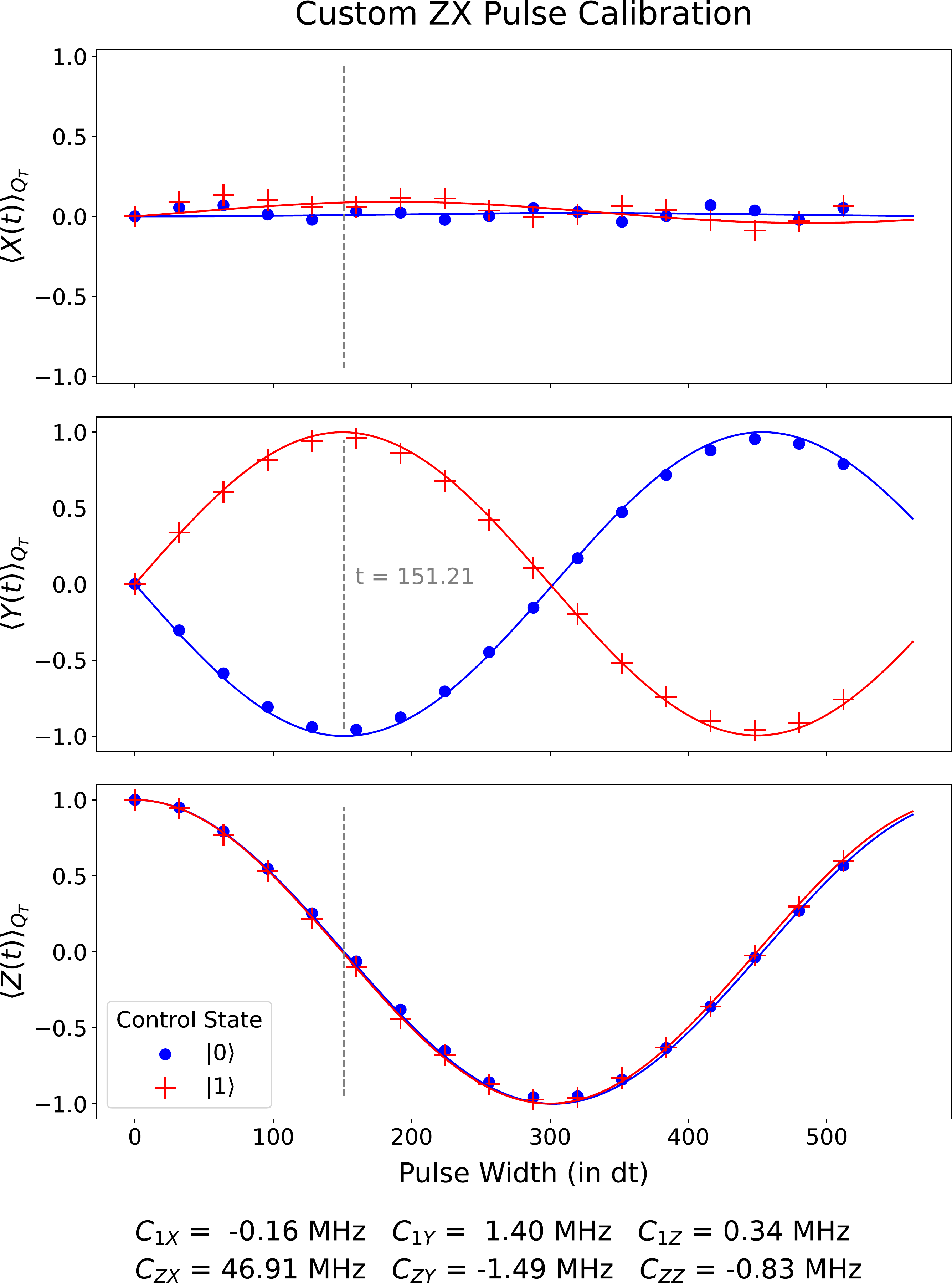}
    \caption{Results of the Hamiltonian tomography experiment using the fully calibrated cross-resonance pulse sequence. We fit the data as described in Ref.~\cite{Hamiltonian_Tomography_2022} to extract the coefficients of the contributions in the interaction Hamiltonian $H_I$ and find the values indicated at the bottom of the figure. The coefficient $C_{ZX}$ of the $Z \otimes X$ term is significantly larger than all other coefficients which indicates a successful calibration. The error bars are smaller than the marker size.}
    \label{fig:full_calibration}
\end{figure}

Third, we need to determine the correct phase and amplitude for the cancellation pulse, which is a resonant flat-top Gaussian pulse on the target qubit with the same duration and Gaussian rise times as the cross-resonance pulse. The purpose of the cancellation pulse is to neutralise the $\mathbf{1} \otimes X$ and $\mathbf{1} \otimes Y$ components in $H_I$. The correct phase for the cancellation tone can be inferred from the Hamiltonian tomography experiment we already performed. By reading off the $C_{\mathbf{1}X}$ and $C_{\mathbf{1}Y}$ coefficients of the $\mathbf{1} \otimes X$ and $\mathbf{1} \otimes Y$ components in $H_I$, we can calculate $\phi_1 = - \tan^{-1}(C_{\mathbf{1}Y} / C_{\mathbf{1}X})$. As the phase of the cross-resonance pulse is set to $\phi_0$, the correct phase for the cancellation tone is $\phi_0 - \phi_1$ as presented in Ref.~\cite{Sheldon_2016}.

To determine the correct amplitude, we perform two Hamiltonian tomography experiments for the full pulse sequence in Fig.~\ref{fig:pulse_schematic}. In the first experiment, the cancellation tone amplitude is set to zero while in the second experiment, we set it to some value $A_0$. The correct order of magnitude for $A_0$ can be estimated from the cancellation tone of the CR($\pi/2$) pulse that forms part of the Controlled-NOT implementation between $Q_C$ and $Q_T$. Hence, we can extract the values $C^1_{\mathbf{1}X}$ and $C^1_{\mathbf{1}Y}$ as well as $C^2_{\mathbf{1}X}$ and $C^2_{\mathbf{1}Y}$ from the two experiments. Assuming a linear relationship between the cancellation tone amplitude and the coefficients as seen in Ref.~\cite{Sheldon_2016}, we find $A_X = A_0 C^1_{\mathbf{1}X}/(C^1_{\mathbf{1}X} - C^2_{\mathbf{1}X})$ and $A_Y = A_0 C^1_{\mathbf{1}Y}/(C^1_{\mathbf{1}Y} - C^2_{\mathbf{1}Y})$.

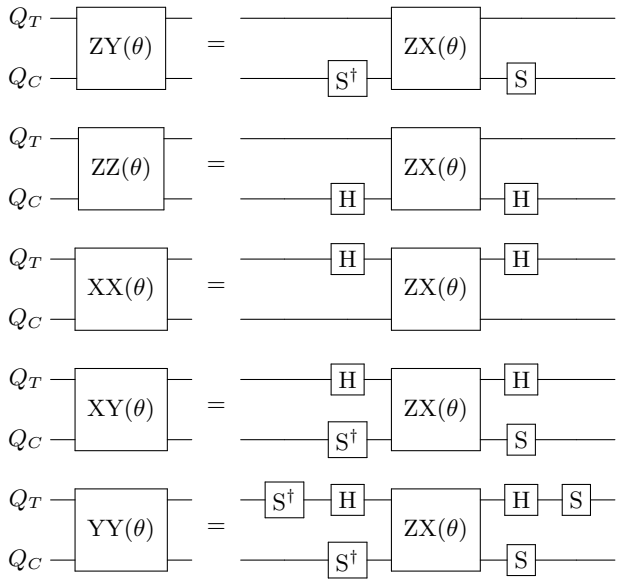
\begin{figure}
    \centering
    \scalebox{1.0}{
    \Qcircuit @C=1.0em @R=1.0em @!R {\\
        Q_T & & 
        \multigate{1}{\mathrm{ZY(\theta)}} & 
        \qw & \raisebox{-2.35em}{=} & & 
        \qw &
        \qw &
        \multigate{1}{\mathrm{ZX(\theta)}} & 
        \qw & 
        \qw &
        \qw
    \\ 
        Q_C & & 
        \ghost{\mathrm{ZY(\theta)}} & 
        \qw & & & 
        \qw &
        \gate{\mathrm{S^\dagger}} & 
        \ghost{\mathrm{ZX(\theta)}} & 
        \gate{\mathrm{S}} & 
        \qw &
        \qw \\
        Q_T & & 
        \multigate{1}{\mathrm{ZZ(\theta)}} & 
        \qw & \raisebox{-2.35em}{=} & & 
        \qw &
        \qw & 
        \multigate{1}{\mathrm{ZX(\theta)}} & 
        \qw & 
        \qw &
        \qw
    \\ 
        Q_C & & 
        \ghost{\mathrm{ZZ(\theta)}} & 
        \qw & & & 
        \qw &
        \gate{\mathrm{H}} & 
        \ghost{\mathrm{ZX(\theta)}} & 
        \gate{\mathrm{H}} & 
        \qw &
        \qw \\
        Q_T & & 
        \multigate{1}{\mathrm{XX(\theta)}} & 
        \qw & \raisebox{-2.35em}{=} & & 
        \qw &
        \gate{\mathrm{H}} & 
        \multigate{1}{\mathrm{ZX(\theta)}} & 
        \gate{\mathrm{H}} & 
        \qw &
        \qw 
    \\
        Q_C & & 
        \ghost{\mathrm{XX(\theta)}} & 
        \qw & & & 
        \qw &
        \qw & 
        \ghost{\mathrm{ZX(\theta)}} & 
        \qw & 
        \qw &
        \qw \\
        Q_T & & 
        \multigate{1}{\mathrm{XY(\theta)}} & 
        \qw & \raisebox{-2.35em}{=} & & 
        \qw & 
        \gate{\mathrm{H}} & 
        \multigate{1}{\mathrm{ZX(\theta)}} & 
        \gate{\mathrm{H}} &
        \qw &
        \qw 
    \\ 
        Q_C & & 
        \ghost{\mathrm{XY(\theta)}} & 
        \qw & & & 
        \qw &
        \gate{\mathrm{S^\dagger}} & 
        \ghost{\mathrm{ZX(\theta)}} & 
        \gate{\mathrm{S}} & 
        \qw &
        \qw\\
        Q_T & & 
        \multigate{1}{\mathrm{YY(\theta)}}& 
        \qw & \raisebox{-2.35em}{=} & & 
        \gate{\mathrm{S^{\dagger}}} & 
        \gate{\mathrm{H}} & 
        \multigate{1}{\mathrm{ZX(\theta)}} & 
        \gate{\mathrm{H}} & 
        \gate{\mathrm{S}} & 
        \qw 
    \\ 
        Q_C & & 
        \ghost{\mathrm{YY(\theta)}} & 
        \qw & & & 
        \qw &
        \gate{\mathrm{S^\dagger}} & 
        \ghost{\mathrm{ZX(\theta)}} & 
        \gate{\mathrm{S}} & 
        \qw &
        \qw \\}
    }
    \caption{Circuit identities that show how the ZX($\theta$) gate can be converted into a range of other two-qubit cross-resonance gates with $H \propto A \otimes B$ where $A,B \in \{X,Y,Z\}$, using single-qubit gates only. The relations are easily shown by applying standard gate identities from Ref.~\cite{Nielsen_Chuang_2010} to the time evolution operator of the ZX($\theta$) gate $U(t) = \cos(gt) \mathbf{1} \otimes \mathbf{1} - i \sin(gt) Z \otimes X$. Here, H indicates the Hadamard gate and S the phase gate.}
    \label{fig:circuit_identities}
\end{figure}

If the value of $\phi_1$ is calibrated correctly, then we find $A_X \approx A_Y$ as the unique solution for the correct amplitude of the cancellation tone \cite{Sheldon_2016}. Hence, to calibrate the full cross-resonance pulse sequence, we only require four Hamiltonian tomography experiments which provides a far more efficient procedure than the calibration methods described in Ref.~\cite{Alexander_2020} and Ref.~\cite{Sheldon_2016}. Furthermore, it is now possible to calibrate the pulse sequence such that it implements a CR($\theta$) gate for values of $\theta$ other than $\pi/2$.

\begin{figure}
    \centering
    \begin{minipage}[t]{0.03\linewidth}
        \vspace{0pt}(a)
    \end{minipage}
    \begin{minipage}[t]{0.95\linewidth}
        \vspace{0pt}
        \includegraphics[width=0.95\linewidth]{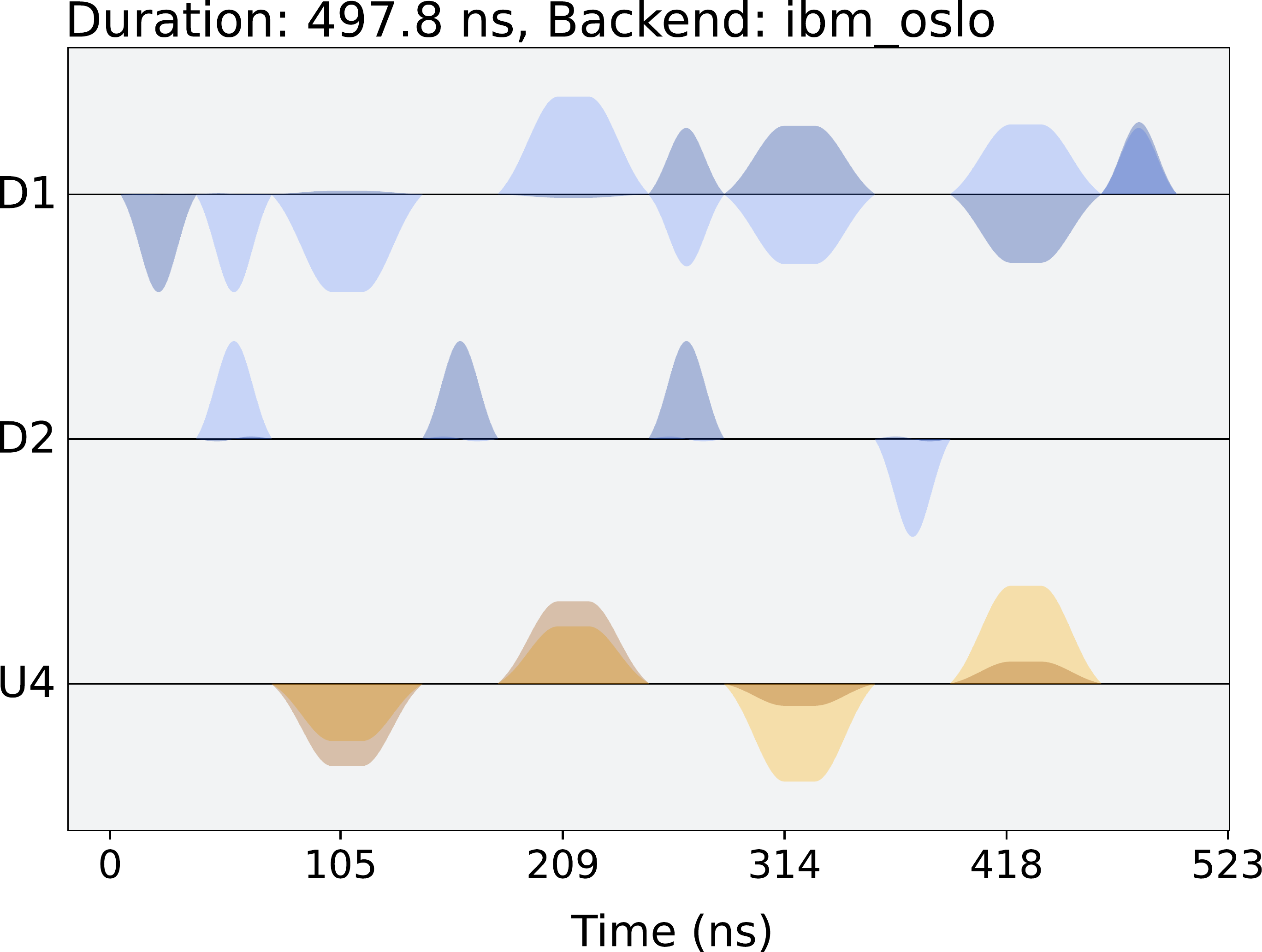}
    \end{minipage}
    \begin{minipage}[t]{0.03\linewidth}
        \vspace{12pt}(b)
    \end{minipage}
    \begin{minipage}[t]{0.95\linewidth}
        \vspace{12pt}
        \includegraphics[width=0.95\linewidth]{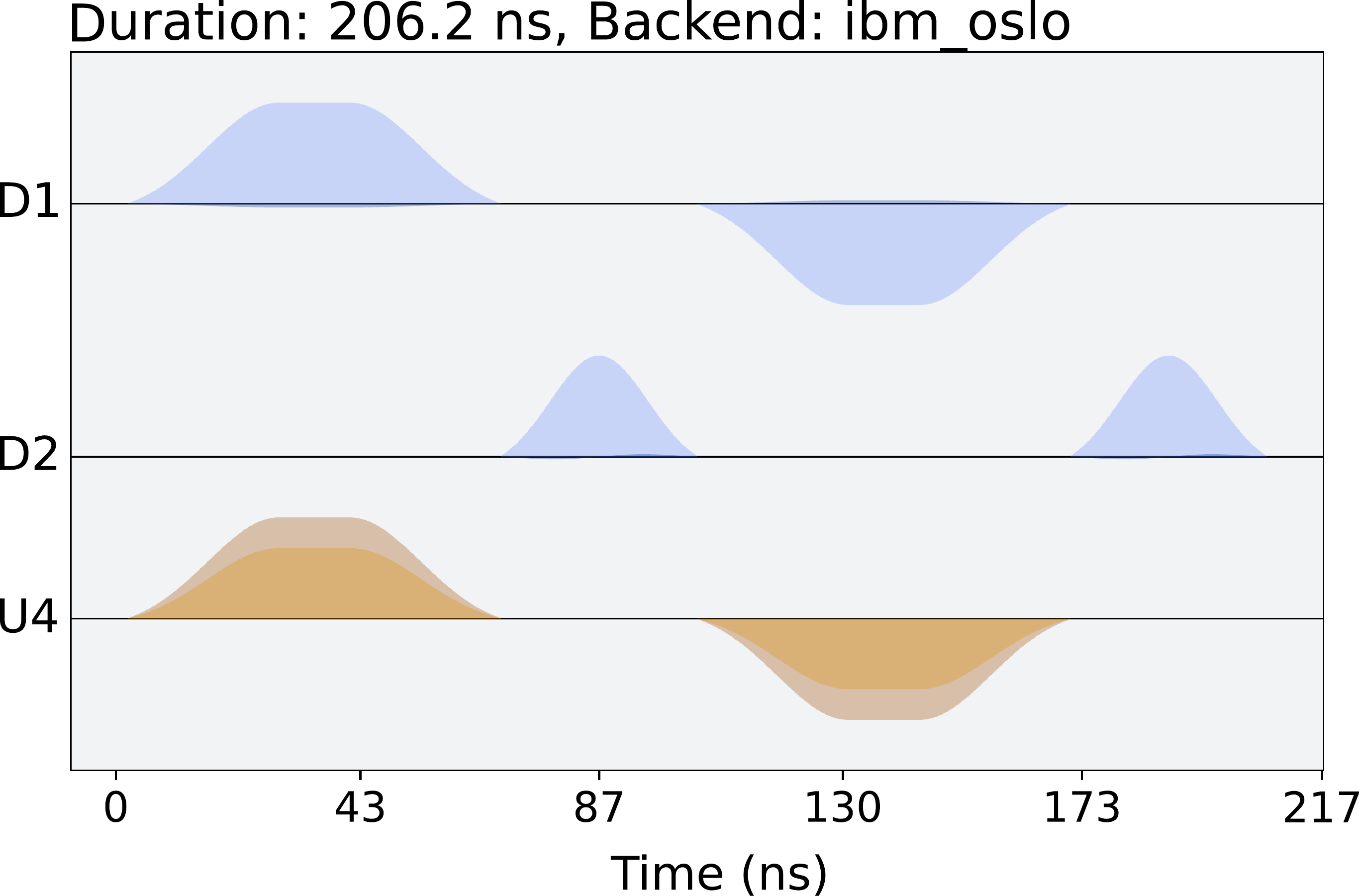}
    \end{minipage}  
    \caption{(a) Schedule for the ZX($\theta$) gate using IBM's circuit-level implementation. The four yellow pulses are the cross-resonance pulses which form part of the two required Controlled-NOT gates. The gate duration is $497.8$ ns. (b) Schedule for the ZX($\theta$) gate using our pulse-level approach which only requires two cross-resonance pulses, halving the number of two-qubit pulses in comparison with the circuit-level approach. The gate duration is reduced to $206.2$ ns.}
    \label{fig:schedule_comparison}
\end{figure}

\begin{figure*}
    \includegraphics[width=\linewidth]{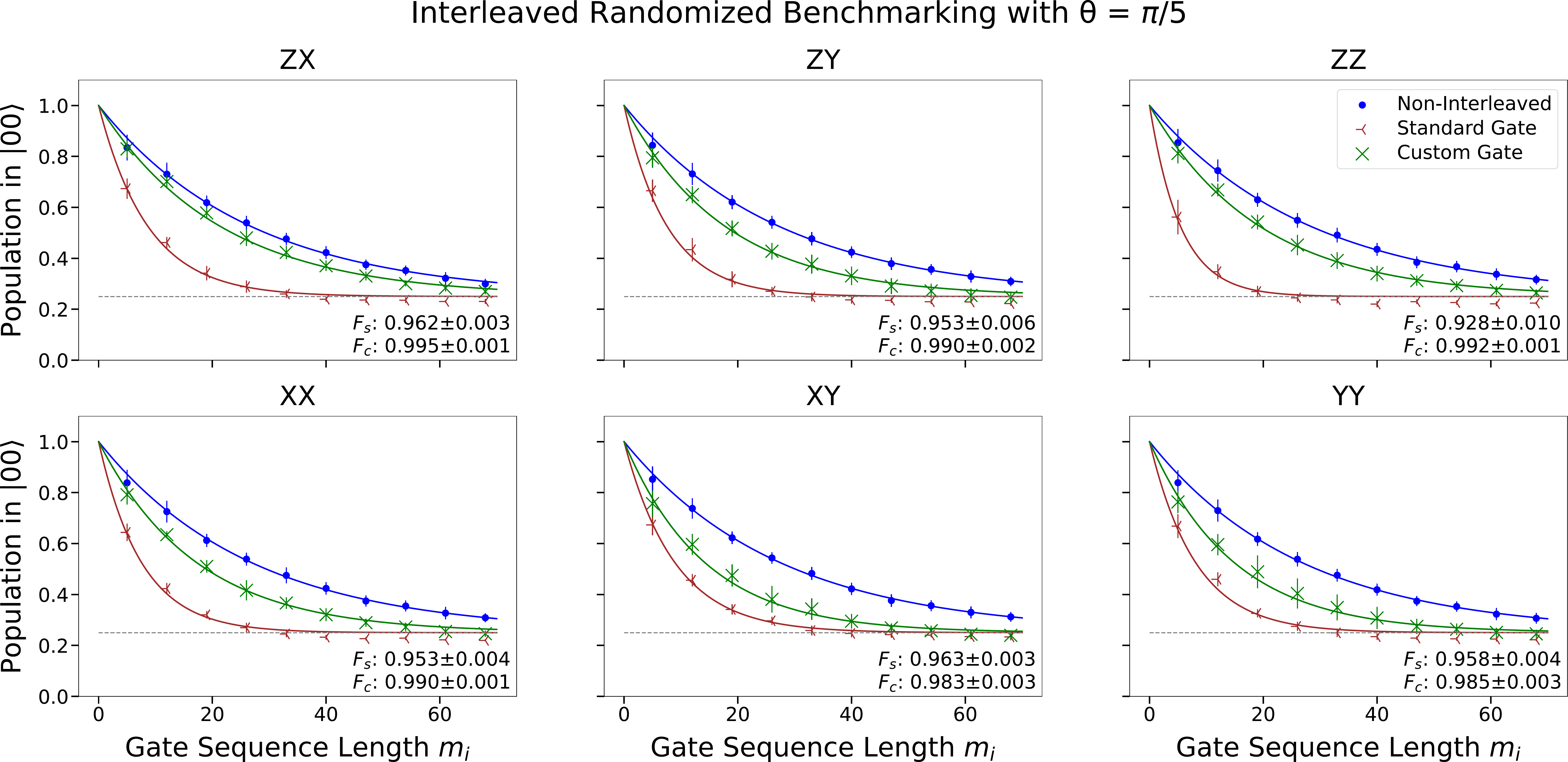}
    \caption{Results of the interleaved randomised benchmarking experiments for six different two-qubit gates in the custom pulse-level and standard circuit-level implementation on a real IBM Quantum backend. Each point indicates the fractional ground state population averaged over ten random gate sequences while error bars show the standard deviations. We observe an exponential decay to the fully mixed state indicated by the dashed line at a fractional ground state population of 0.25. For each gate, $F_S$ and $F_C$ characterise the respective noise resistances of the standard and custom implementation as discussed in the main text. In all cases, we find that $F_C \ge F_S$ marks a significant improvement in the noise resistances of the gates.}
    \label{fig:irb_results}
\end{figure*}

We have implemented this calibration procedure using the seven-qubit IBM Quantum backend \verb|ibm_oslo| with qubit 2 and qubit 1 as the control qubit and target qubit, respectively. The resonance frequency and anharmonicity of the control qubit are $f_2 = 4.962$ GHz and $\delta_2 = -0.344$ GHz, and $f_1 = 5.046$ GHz and $\delta_1 = -0.343$ GHz for the target qubit \cite{ibm_resources}. To illustrate our generalised procedure by way of example, we calibrate a CR($\theta$) gate for $\theta = \pi/5$. In all calibration experiments, we use at least $4~000$ repetitions per circuit such that the statistical errors become negligibly small. We also mitigate readout errors using the method described in Ref.~\cite{Readout_Error}. The results of the amplitude calibration experiment are illustrated in Fig.~\ref{fig:amplitude_calibration} and the results of an experiment that verifies the calibration are displayed in Fig.~\ref{fig:full_calibration}. For these two experiments, we have used $20~000$ repetitions per circuit. Setting the pulse width to the inherited width as described above, we receive the CR($\pi/5$) gate. This demonstrates that we can use our runtime-efficient, pulse-level procedure to calibrate a CR($\theta$) gate for $\theta$ other than $\pi/2$.

Having implemented the CR($\theta$) gate with $H \propto Z \otimes X$, it is straightforward to implement a range of other two-qubit interactions. As illustrated in Fig.~\ref{fig:circuit_identities}, we can use standard single-qubit gates on $Q_T$ and $Q_C$ to convert the $Z \otimes X$ interaction into any $A \otimes B$ interaction with $A,B \in \{X, Y, Z\}$. Note that we have written the gate that corresponds to the Hamiltonian $H = (-\theta/2) A \otimes B$ as AB($\theta$) for ease of notation. The relations in Fig.~\ref{fig:circuit_identities} are easily proven using standard gate identities \cite{Nielsen_Chuang_2010}. Finally, the XZ($\theta$), YZ($\theta$), and YX($\theta$) gates can either be implemented by circuit identities used in Fig.~\ref{fig:circuit_identities}, or alternatively by swapping the control and target qubit in the calibration procedure. Hence, we conclude that having calibrated the ZX($\theta$) gate, it is straightforward to implement any of the nine AB($\theta$) gates with $A,B \in \{X, Y, Z\}$.

Using pulse-level methods, we can further extend the set of easily implemented two-qubit gates. Note that the $S$ and $S^\dagger$ gates in Fig.~\ref{fig:circuit_identities} correspond to virtual phase shifts with $\Delta\phi = \pm\pi/2$ on the relevant qubit, respectively \cite{McKay_2017}. As Qiskit Pulse allows us to directly specify a phase shift, we can also implement values of $\Delta\phi$ other than $\pi/2$. For instance, by shifting the phase of the cross-resonance pulse and cancellation tone by $\Delta\phi_0$, we can convert the ZX($\theta$) gate into a Z($\cos(\Delta\phi_0)$X+$\sin(\Delta\phi_0)$Y)($\theta$) gate. While this treatment is not exhaustive, it nicely illustrates that using circuit-level and pulse-level methods, a range of two-qubit interactions are straightforwardly implemented once we have calibrated the ZX($\theta$) gate.

Any of these gates can also be implemented using circuit-level methods with at most three Controlled-NOT gates \cite{Vatan_2004}. However, the advantage of our pulse-level implementation of the ZX($\theta$) gate is that we require fewer cross-resonance pulses as illustrated in Fig.~\ref{fig:schedule_comparison}. As two-qubit interactions on superconducting qubits are susceptible to noise \cite{Krantz_2019}, minimising the number of cross-resonance pulses should increase the noise resistance of the ZX($\theta$) gate. Further, as single-qubit gates achieve near-perfect fidelities \cite{Sheldon_2016(2)} while virtual phase gates have perfect fidelities \cite{McKay_2017}, converting our ZX($\theta$) gate into other two-qubit gates as in Fig.~\ref{fig:circuit_identities} should lead to similar improvements for the noise resistances of these gates.

We test both hypotheses by performing an interleaved randomised benchmarking experiment similar to those in Ref.~\cite{Magesan_2012} and Ref.~\cite{Interleaved_RB}. To measure the noise resistance we proceed as follows. We define a set of gate sequence lengths $\{m_1, ..., m_j\}$ with $\Delta = m_{i+1} - m_i$ some fixed positive integer and $m_j = N$. Using the \verb|StandardRB| method in Qiskit Pulse, we sample a set of random gate sequences $\{C_1, ..., C_N\}$ and define a new set of gate sequences $\mathbf{R} = \{R_{m_1}, ..., R_{m_j}\}$ with $R_{m_i} = C_1...C_{m_i}\tilde{C}_{m_i}$ where $\tilde{C}_{m_i}$ inverts the previous operations such that the action of each $R_{m_i}$ is just the identity operation. Then, to measure the noise resistance of the standard and custom ZX($\theta$) gate, we interleave ZX($\theta$)ZX($-\theta$) after every $C_l$ in each $R_{m_i}$, giving two sets of gate sequences $\mathbf{R_S}$ and $\mathbf{R_C}$, respectively. We run each of the gate sequences in $\mathbf{R}$, $\mathbf{R_S}$ and $\mathbf{R_C}$, and measure the fractional ground state population. As the sequences in $\mathbf{R_S}$ and $\mathbf{R_C}$ are identity operations that acquire an additional error due to the interleaved cross-resonance gates, we expect the ground state population to decay faster by a factor of $F^{2m_i}$, where $F \leq 1$ characterises the additional error introduced by the ZX($\theta$) gate, in comparison to the case of non-interleaved gate sequences in $\mathbf{R}$. By fitting the data to an exponential decay, we find the values of $F_S$ and $F_C$.

We have implemented the interleaved benchmarking procedure using the same quantum backend and qubits as described above. The gates ZX, ZY, ZZ, XY, XX and YY are tested using $m_1 = 5$, $\Delta = 7$, and $N = 68$ as benchmarking parameters. For the custom implementation we have used the ZX($\pi/5$) gate implemented above while for the standard implementation we have used circuit-level methods in Qiskit \cite{Qiskit_Gates}, employing the circuit identities in Fig.~\ref{fig:circuit_identities} where necessary. For each gate, we repeat the experiment with ten random gate sequences using $20~000$ repetitions per circuit. The results are shown in Fig.~\ref{fig:irb_results}.

We must be careful in interpreting $F_S$ and $F_C$ as they do not characterise the total gate error but rather the error associated with performing an identity operation by using the ZX($\theta$) gate and its inverse. In general, we expect this error to come from coherent errors and noise. Performing additional Hamiltonian tomography experiments for the ZX($\theta$) gate and its inverse both in the custom and standard implementation, we measure similar coefficients for all terms in the respective Hamiltonians. This allows us to rule out the possibility that the large discrepancy between $F_S$ and $F_C$ is due to coherent errors.

Hence, we can interpret $F_S$ and $F_C$ as characterising the error from noise for the standard and custom ZX$(\theta)$ gate implementations, respectively. With this interpretation, we can verify both hypotheses. First, for the ZX($\theta$) gate, we observe that $F_C$ is significantly larger than $F_S$ in Fig.~\ref{fig:irb_results}, indicating that our custom implementation, requiring fewer cross-resonance pulses, is more resistant to noise. And second, we see similar improvements for the noise resistances of the other gates that we tested in Fig.~\ref{fig:irb_results}, as expected due to high single-qubit gate fidelities. Therefore, our pulse-level implementation of the ZX($\theta$) gate provides us with a wide range of two-qubit gates with significantly improved noise resistances. Since the overall pulse schedule time is significantly shorter than the coherence time of the qubits \cite{ibm_resources}, we conjecture that this improvement is due to the simplified pulse architecture we developed, rather than the reduced gate time.

Finally, we comment on the relevance of this result for practical quantum computing. While generally advantageous, improved noise resistances for two-qubit operations are particularly useful in Hamiltonian Simulation. This often requires the repeated application of multi-qubit gates, for instance in Trotterisation approaches \cite{Nielsen_Chuang_2010}, and thus remains limited by the low noise resistances of multi-qubit interactions. By calibrating a gate using our pulse-level approach, the noise resistance can be significantly improved. This can enable Hamiltonian Simulation, as we will present in a subsequent paper \cite{Tennie_2023}. In this sense, our procedure is not just interesting from an engineering but also from a physics perspective, as we can use the improved gates to simulate interesting physical systems on publicly available IBM quantum backends.

To conclude, we provide a powerful extension to the set of high-fidelity, multi-qubit gates on currently available quantum computers based on superconducting qubits. With our runtime-efficient and reproducible pulse-level approach, one can calibrate a CR($\theta$) cross-resonance gate for a given value of $\theta$ which is extended to a wide range of other two-qubit gates by applying single-qubit gates. We have demonstrated that this pulse-level approach, requiring fewer two-qubit pulses than the circuit-level approach currently used by IBM, significantly improves the noise resistances of the CR($\theta$) gate and related interactions.

While providing a compelling proof of principle, we were limited to performing experiments on publicly available IBM Quantum backends. Future work should focus on repeating our experiments on those IBM Quantum backends which are currently not available to the general public. Further, our pulse-level approach should be tested in quantum computing applications to demonstrate the practical usefulness of the improvement. This will be explored for Hamiltonian Simulation in a subsequent paper.

\begin{acknowledgments}
We acknowledge the use of IBM Quantum services for this work. The views expressed are those of the authors, and do not reflect the official policy or position of IBM or the IBM Quantum team. 
DD acknowledges support from the Studienstiftung des Deutschen Volkes. FT acknowledges support from the UKRI New Horizons Grant EP/X017249/1.
\end{acknowledgments}

\bibliographystyle{apsrev4-1}
\bibliography{working_doc}

\begin{thebibliography}{23}%
\makeatletter
\providecommand \@ifxundefined [1]{%
 \@ifx{#1\undefined}
}%
\providecommand \@ifnum [1]{%
 \ifnum #1\expandafter \@firstoftwo
 \else \expandafter \@secondoftwo
 \fi
}%
\providecommand \@ifx [1]{%
 \ifx #1\expandafter \@firstoftwo
 \else \expandafter \@secondoftwo
 \fi
}%
\providecommand \natexlab [1]{#1}%
\providecommand \enquote  [1]{``#1''}%
\providecommand \bibnamefont  [1]{#1}%
\providecommand \bibfnamefont [1]{#1}%
\providecommand \citenamefont [1]{#1}%
\providecommand \href@noop [0]{\@secondoftwo}%
\providecommand \href [0]{\begingroup \@sanitize@url \@href}%
\providecommand \@href[1]{\@@startlink{#1}\@@href}%
\providecommand \@@href[1]{\endgroup#1\@@endlink}%
\providecommand \@sanitize@url [0]{\catcode `\\12\catcode `\$12\catcode
  `\&12\catcode `\#12\catcode `\^12\catcode `\_12\catcode `\%12\relax}%
\providecommand \@@startlink[1]{}%
\providecommand \@@endlink[0]{}%
\providecommand \url  [0]{\begingroup\@sanitize@url \@url }%
\providecommand \@url [1]{\endgroup\@href {#1}{\urlprefix }}%
\providecommand \urlprefix  [0]{URL }%
\providecommand \Eprint [0]{\href }%
\providecommand \doibase [0]{http://dx.doi.org/}%
\providecommand \selectlanguage [0]{\@gobble}%
\providecommand \bibinfo  [0]{\@secondoftwo}%
\providecommand \bibfield  [0]{\@secondoftwo}%
\providecommand \translation [1]{[#1]}%
\providecommand \BibitemOpen [0]{}%
\providecommand \bibitemStop [0]{}%
\providecommand \bibitemNoStop [0]{.\EOS\space}%
\providecommand \EOS [0]{\spacefactor3000\relax}%
\providecommand \BibitemShut  [1]{\csname bibitem#1\endcsname}%
\let\auto@bib@innerbib\@empty
\bibitem [{\citenamefont {Montanaro}(2016)}]{Montanaro_2016}%
  \BibitemOpen
  \bibfield  {author} {\bibinfo {author} {\bibfnamefont {A.}~\bibnamefont
  {Montanaro}},\ }\href {\doibase 10.1038/npjqi.2015.23} {\bibfield  {journal}
  {\bibinfo  {journal} {npj Quantum Inf}\ }\textbf {\bibinfo {volume} {2}},\
  \bibinfo {pages} {15023} (\bibinfo {year} {2016})}\BibitemShut {NoStop}%
\bibitem [{\citenamefont {Nielsen}\ and\ \citenamefont
  {Chuang}(2010)}]{Nielsen_Chuang_2010}%
  \BibitemOpen
  \bibfield  {author} {\bibinfo {author} {\bibfnamefont {M.~A.}\ \bibnamefont
  {Nielsen}}\ and\ \bibinfo {author} {\bibfnamefont {I.~L.}\ \bibnamefont
  {Chuang}},\ }\href {\doibase 10.1017/CBO9780511976667} {\emph {\bibinfo
  {title} {Quantum Computation and Quantum Information: 10th Anniversary
  Edition}}}\ (\bibinfo  {publisher} {Cambridge University Press},\ \bibinfo
  {year} {2010})\BibitemShut {NoStop}%
\bibitem [{\citenamefont {Brooks}(2023)}]{Brooks_2023}%
  \BibitemOpen
  \bibfield  {author} {\bibinfo {author} {\bibfnamefont {M.}~\bibnamefont
  {Brooks}},\ }\href@noop {} {\enquote {\bibinfo {title} {What’s next for
  quantum computing},}\ } (\bibinfo {year} {2023}),\ \bibinfo {note} {{MIT
  Technology Review}
  \url{https://www.technologyreview.com/2023/01/06/1066317/whats-next-for-quantum-computing/}}\BibitemShut
  {NoStop}%
\bibitem [{\citenamefont {Kjaergaard}\ \emph {et~al.}(2020)\citenamefont
  {Kjaergaard}, \citenamefont {Schwartz}, \citenamefont {Braumüller},
  \citenamefont {Krantz}, \citenamefont {Wang}, \citenamefont {Gustavsson},\
  and\ \citenamefont {Oliver}}]{Kjaergaard_2020}%
  \BibitemOpen
  \bibfield  {author} {\bibinfo {author} {\bibfnamefont {M.}~\bibnamefont
  {Kjaergaard}}, \bibinfo {author} {\bibfnamefont {M.~E.}\ \bibnamefont
  {Schwartz}}, \bibinfo {author} {\bibfnamefont {J.}~\bibnamefont
  {Braumüller}}, \bibinfo {author} {\bibfnamefont {P.}~\bibnamefont {Krantz}},
  \bibinfo {author} {\bibfnamefont {J.~I.-J.}\ \bibnamefont {Wang}}, \bibinfo
  {author} {\bibfnamefont {S.}~\bibnamefont {Gustavsson}}, \ and\ \bibinfo
  {author} {\bibfnamefont {W.~D.}\ \bibnamefont {Oliver}},\ }\href {\doibase
  10.1146/annurev-conmatphys-031119-050605} {\bibfield  {journal} {\bibinfo
  {journal} {Annu. Rev. Condens. Matter Phys.}\ }\textbf {\bibinfo {volume}
  {11}},\ \bibinfo {pages} {369} (\bibinfo {year} {2020})}\BibitemShut
  {NoStop}%
\bibitem [{\citenamefont {Glaser}\ \emph {et~al.}(2015)\citenamefont {Glaser},
  \citenamefont {Boscain}, \citenamefont {Calarco}, \citenamefont {Koch},
  \citenamefont {K{\"o}ckenberger}, \citenamefont {Kosloff}, \citenamefont
  {Kuprov}, \citenamefont {Luy}, \citenamefont {Schirmer}, \citenamefont
  {Schulte-Herbr{\"u}ggen}, \citenamefont {Sugny},\ and\ \citenamefont
  {Wilhelm}}]{Glaser_2015}%
  \BibitemOpen
  \bibfield  {author} {\bibinfo {author} {\bibfnamefont {S.~J.}\ \bibnamefont
  {Glaser}}, \bibinfo {author} {\bibfnamefont {U.}~\bibnamefont {Boscain}},
  \bibinfo {author} {\bibfnamefont {T.}~\bibnamefont {Calarco}}, \bibinfo
  {author} {\bibfnamefont {C.~P.}\ \bibnamefont {Koch}}, \bibinfo {author}
  {\bibfnamefont {W.}~\bibnamefont {K{\"o}ckenberger}}, \bibinfo {author}
  {\bibfnamefont {R.}~\bibnamefont {Kosloff}}, \bibinfo {author} {\bibfnamefont
  {I.}~\bibnamefont {Kuprov}}, \bibinfo {author} {\bibfnamefont
  {B.}~\bibnamefont {Luy}}, \bibinfo {author} {\bibfnamefont {S.}~\bibnamefont
  {Schirmer}}, \bibinfo {author} {\bibfnamefont {T.}~\bibnamefont
  {Schulte-Herbr{\"u}ggen}}, \bibinfo {author} {\bibfnamefont {D.}~\bibnamefont
  {Sugny}}, \ and\ \bibinfo {author} {\bibfnamefont {F.~K.}\ \bibnamefont
  {Wilhelm}},\ }\href {\doibase 10.1140/epjd/e2015-60464-1} {\bibfield
  {journal} {\bibinfo  {journal} {Eur. Phys. J. D}\ }\textbf {\bibinfo {volume}
  {69}},\ \bibinfo {pages} {279} (\bibinfo {year} {2015})}\BibitemShut
  {NoStop}%
\bibitem [{Qis(2021)}]{Qiskit}%
  \BibitemOpen
  \href {\doibase 10.5281/zenodo.2573505} {\enquote {\bibinfo {title} {Qiskit:
  An open-source framework for quantum computing},}\ } (\bibinfo {year}
  {2021})\BibitemShut {NoStop}%
\bibitem [{IBM()}]{IBMQuantumLab}%
  \BibitemOpen
  \href@noop {} {}\bibinfo {note} {{"IBM Quantum,"},
  \url{https://quantum-computing.ibm.com} (2023)}\BibitemShut {NoStop}%
\bibitem [{\citenamefont {Alexander}\ \emph {et~al.}(2020)\citenamefont
  {Alexander}, \citenamefont {Kanazawa}, \citenamefont {Egger}, \citenamefont
  {Capelluto}, \citenamefont {Wood}, \citenamefont {Javadi-Abhari},\ and\
  \citenamefont {McKay}}]{Alexander_2020}%
  \BibitemOpen
  \bibfield  {author} {\bibinfo {author} {\bibfnamefont {T.}~\bibnamefont
  {Alexander}}, \bibinfo {author} {\bibfnamefont {N.}~\bibnamefont {Kanazawa}},
  \bibinfo {author} {\bibfnamefont {D.~J.}\ \bibnamefont {Egger}}, \bibinfo
  {author} {\bibfnamefont {L.}~\bibnamefont {Capelluto}}, \bibinfo {author}
  {\bibfnamefont {C.~J.}\ \bibnamefont {Wood}}, \bibinfo {author}
  {\bibfnamefont {A.}~\bibnamefont {Javadi-Abhari}}, \ and\ \bibinfo {author}
  {\bibfnamefont {D.~C.}\ \bibnamefont {McKay}},\ }\href {\doibase
  10.1088/2058-9565/aba404} {\bibfield  {journal} {\bibinfo  {journal} {Quantum
  Sci. Technol.}\ }\textbf {\bibinfo {volume} {5}},\ \bibinfo {pages} {044006}
  (\bibinfo {year} {2020})}\BibitemShut {NoStop}%
\bibitem [{\citenamefont {Krantz}\ \emph {et~al.}(2019)\citenamefont {Krantz},
  \citenamefont {Kjaergaard}, \citenamefont {Yan}, \citenamefont {Orlando},
  \citenamefont {Gustavsson},\ and\ \citenamefont {Oliver}}]{Krantz_2019}%
  \BibitemOpen
  \bibfield  {author} {\bibinfo {author} {\bibfnamefont {P.}~\bibnamefont
  {Krantz}}, \bibinfo {author} {\bibfnamefont {M.}~\bibnamefont {Kjaergaard}},
  \bibinfo {author} {\bibfnamefont {F.}~\bibnamefont {Yan}}, \bibinfo {author}
  {\bibfnamefont {T.~P.}\ \bibnamefont {Orlando}}, \bibinfo {author}
  {\bibfnamefont {S.}~\bibnamefont {Gustavsson}}, \ and\ \bibinfo {author}
  {\bibfnamefont {W.~D.}\ \bibnamefont {Oliver}},\ }\href {\doibase
  10.1063/1.5089550} {\bibfield  {journal} {\bibinfo  {journal} {Appl. Phys.
  Rev.}\ }\textbf {\bibinfo {volume} {6}},\ \bibinfo {pages} {021318} (\bibinfo
  {year} {2019})}\BibitemShut {NoStop}%
\bibitem [{\citenamefont {Gambetta}\ \emph {et~al.}(2017)\citenamefont
  {Gambetta}, \citenamefont {Chow},\ and\ \citenamefont
  {Steffen}}]{Gambetta_2017}%
  \BibitemOpen
  \bibfield  {author} {\bibinfo {author} {\bibfnamefont {J.~M.}\ \bibnamefont
  {Gambetta}}, \bibinfo {author} {\bibfnamefont {J.~M.}\ \bibnamefont {Chow}},
  \ and\ \bibinfo {author} {\bibfnamefont {M.}~\bibnamefont {Steffen}},\ }\href
  {\doibase 10.1038/s41534-016-0004-0} {\bibfield  {journal} {\bibinfo
  {journal} {npj Quantum Information}\ }\textbf {\bibinfo {volume} {3}},\
  \bibinfo {pages} {2} (\bibinfo {year} {2017})}\BibitemShut {NoStop}%
\bibitem [{\citenamefont {Rigetti}\ and\ \citenamefont
  {Devoret}(2010)}]{Rigetti_2010}%
  \BibitemOpen
  \bibfield  {author} {\bibinfo {author} {\bibfnamefont {C.}~\bibnamefont
  {Rigetti}}\ and\ \bibinfo {author} {\bibfnamefont {M.}~\bibnamefont
  {Devoret}},\ }\href {\doibase 10.1103/PhysRevB.81.134507} {\bibfield
  {journal} {\bibinfo  {journal} {Phys. Rev. B}\ }\textbf {\bibinfo {volume}
  {81}},\ \bibinfo {pages} {134507} (\bibinfo {year} {2010})}\BibitemShut
  {NoStop}%
\bibitem [{\citenamefont {Sheldon}\ \emph
  {et~al.}(2016{\natexlab{a}})\citenamefont {Sheldon}, \citenamefont {Magesan},
  \citenamefont {Chow},\ and\ \citenamefont {Gambetta}}]{Sheldon_2016}%
  \BibitemOpen
  \bibfield  {author} {\bibinfo {author} {\bibfnamefont {S.}~\bibnamefont
  {Sheldon}}, \bibinfo {author} {\bibfnamefont {E.}~\bibnamefont {Magesan}},
  \bibinfo {author} {\bibfnamefont {J.~M.}\ \bibnamefont {Chow}}, \ and\
  \bibinfo {author} {\bibfnamefont {J.~M.}\ \bibnamefont {Gambetta}},\ }\href
  {\doibase 10.1103/PhysRevA.93.060302} {\bibfield  {journal} {\bibinfo
  {journal} {Phys. Rev. A}\ }\textbf {\bibinfo {volume} {93}},\ \bibinfo
  {pages} {060302} (\bibinfo {year} {2016}{\natexlab{a}})}\BibitemShut
  {NoStop}%
\bibitem [{\citenamefont {Magesan}\ and\ \citenamefont
  {Gambetta}(2020)}]{Magesan_2020}%
  \BibitemOpen
  \bibfield  {author} {\bibinfo {author} {\bibfnamefont {E.}~\bibnamefont
  {Magesan}}\ and\ \bibinfo {author} {\bibfnamefont {J.~M.}\ \bibnamefont
  {Gambetta}},\ }\href {\doibase 10.1103/PhysRevA.101.052308} {\bibfield
  {journal} {\bibinfo  {journal} {Phys. Rev. A}\ }\textbf {\bibinfo {volume}
  {101}},\ \bibinfo {pages} {052308} (\bibinfo {year} {2020})}\BibitemShut
  {NoStop}%
\bibitem [{\citenamefont {Magesan}\ \emph {et~al.}(2012)\citenamefont
  {Magesan}, \citenamefont {Gambetta}, \citenamefont {Johnson}, \citenamefont
  {Ryan}, \citenamefont {Chow}, \citenamefont {Merkel}, \citenamefont
  {da~Silva}, \citenamefont {Keefe}, \citenamefont {Rothwell}, \citenamefont
  {Ohki}, \citenamefont {Ketchen},\ and\ \citenamefont
  {Steffen}}]{Magesan_2012}%
  \BibitemOpen
  \bibfield  {author} {\bibinfo {author} {\bibfnamefont {E.}~\bibnamefont
  {Magesan}}, \bibinfo {author} {\bibfnamefont {J.~M.}\ \bibnamefont
  {Gambetta}}, \bibinfo {author} {\bibfnamefont {B.~R.}\ \bibnamefont
  {Johnson}}, \bibinfo {author} {\bibfnamefont {C.~A.}\ \bibnamefont {Ryan}},
  \bibinfo {author} {\bibfnamefont {J.~M.}\ \bibnamefont {Chow}}, \bibinfo
  {author} {\bibfnamefont {S.~T.}\ \bibnamefont {Merkel}}, \bibinfo {author}
  {\bibfnamefont {M.~P.}\ \bibnamefont {da~Silva}}, \bibinfo {author}
  {\bibfnamefont {G.~A.}\ \bibnamefont {Keefe}}, \bibinfo {author}
  {\bibfnamefont {M.~B.}\ \bibnamefont {Rothwell}}, \bibinfo {author}
  {\bibfnamefont {T.~A.}\ \bibnamefont {Ohki}}, \bibinfo {author}
  {\bibfnamefont {M.~B.}\ \bibnamefont {Ketchen}}, \ and\ \bibinfo {author}
  {\bibfnamefont {M.}~\bibnamefont {Steffen}},\ }\href {\doibase
  10.1103/PhysRevLett.109.080505} {\bibfield  {journal} {\bibinfo  {journal}
  {Phys. Rev. Lett.}\ }\textbf {\bibinfo {volume} {109}},\ \bibinfo {pages}
  {080505} (\bibinfo {year} {2012})}\BibitemShut {NoStop}%
\bibitem [{\citenamefont {{The Qiskit
  Team}}(2022{\natexlab{a}})}]{Hamiltonian_Tomography_2022}%
  \BibitemOpen
  \bibfield  {author} {\bibinfo {author} {\bibnamefont {{The Qiskit Team}}},\
  }\href@noop {} {\enquote {\bibinfo {title} {{Hamiltonian Tomography}},}\
  }\bibinfo {howpublished}
  {\url{https://qiskit.org/textbook/ch-quantum-hardware/hamiltonian-tomography.html}}
  (\bibinfo {year} {2022}{\natexlab{a}})\BibitemShut {NoStop}%
\bibitem [{ibm()}]{ibm_resources}%
  \BibitemOpen
  \href@noop {} {}\bibinfo {note} {{"IBM Quantum Resources,"},
  \url{https://quantum-computing.ibm.com/services/resources}
  (2023)}\BibitemShut {NoStop}%
\bibitem [{\citenamefont {{The Qiskit
  Team}}(2022{\natexlab{b}})}]{Readout_Error}%
  \BibitemOpen
  \bibfield  {author} {\bibinfo {author} {\bibnamefont {{The Qiskit Team}}},\
  }\href@noop {} {\enquote {\bibinfo {title} {{Readout Mitigation}},}\
  }\bibinfo {howpublished}
  {\url{https://qiskit.org/documentation/experiments/tutorials/readout_mitigation.html}}
  (\bibinfo {year} {2022}{\natexlab{b}})\BibitemShut {NoStop}%
\bibitem [{\citenamefont {McKay}\ \emph {et~al.}(2017)\citenamefont {McKay},
  \citenamefont {Wood}, \citenamefont {Sheldon}, \citenamefont {Chow},\ and\
  \citenamefont {Gambetta}}]{McKay_2017}%
  \BibitemOpen
  \bibfield  {author} {\bibinfo {author} {\bibfnamefont {D.~C.}\ \bibnamefont
  {McKay}}, \bibinfo {author} {\bibfnamefont {C.~J.}\ \bibnamefont {Wood}},
  \bibinfo {author} {\bibfnamefont {S.}~\bibnamefont {Sheldon}}, \bibinfo
  {author} {\bibfnamefont {J.~M.}\ \bibnamefont {Chow}}, \ and\ \bibinfo
  {author} {\bibfnamefont {J.~M.}\ \bibnamefont {Gambetta}},\ }\href {\doibase
  10.1103/PhysRevA.96.022330} {\bibfield  {journal} {\bibinfo  {journal} {Phys.
  Rev. A}\ }\textbf {\bibinfo {volume} {96}},\ \bibinfo {pages} {022330}
  (\bibinfo {year} {2017})}\BibitemShut {NoStop}%
\bibitem [{\citenamefont {Vatan}\ and\ \citenamefont
  {Williams}(2004)}]{Vatan_2004}%
  \BibitemOpen
  \bibfield  {author} {\bibinfo {author} {\bibfnamefont {F.}~\bibnamefont
  {Vatan}}\ and\ \bibinfo {author} {\bibfnamefont {C.}~\bibnamefont
  {Williams}},\ }\href {\doibase 10.1103/PhysRevA.69.032315} {\bibfield
  {journal} {\bibinfo  {journal} {Phys. Rev. A}\ }\textbf {\bibinfo {volume}
  {69}},\ \bibinfo {pages} {032315} (\bibinfo {year} {2004})}\BibitemShut
  {NoStop}%
\bibitem [{\citenamefont {Sheldon}\ \emph
  {et~al.}(2016{\natexlab{b}})\citenamefont {Sheldon}, \citenamefont {Bishop},
  \citenamefont {Magesan}, \citenamefont {Filipp}, \citenamefont {Chow},\ and\
  \citenamefont {Gambetta}}]{Sheldon_2016(2)}%
  \BibitemOpen
  \bibfield  {author} {\bibinfo {author} {\bibfnamefont {S.}~\bibnamefont
  {Sheldon}}, \bibinfo {author} {\bibfnamefont {L.~S.}\ \bibnamefont {Bishop}},
  \bibinfo {author} {\bibfnamefont {E.}~\bibnamefont {Magesan}}, \bibinfo
  {author} {\bibfnamefont {S.}~\bibnamefont {Filipp}}, \bibinfo {author}
  {\bibfnamefont {J.~M.}\ \bibnamefont {Chow}}, \ and\ \bibinfo {author}
  {\bibfnamefont {J.~M.}\ \bibnamefont {Gambetta}},\ }\href {\doibase
  10.1103/PhysRevA.93.012301} {\bibfield  {journal} {\bibinfo  {journal} {Phys.
  Rev. A}\ }\textbf {\bibinfo {volume} {93}},\ \bibinfo {pages} {012301}
  (\bibinfo {year} {2016}{\natexlab{b}})}\BibitemShut {NoStop}%
\bibitem [{\citenamefont {{The Qiskit
  Team}}(2022{\natexlab{c}})}]{Interleaved_RB}%
  \BibitemOpen
  \bibfield  {author} {\bibinfo {author} {\bibnamefont {{The Qiskit Team}}},\
  }\href@noop {} {\enquote {\bibinfo {title} {{Randomized Benchmarking}},}\
  }\bibinfo {howpublished}
  {\url{https://qiskit.org/documentation/experiments/tutorials/randomized_benchmarking.html}}
  (\bibinfo {year} {2022}{\natexlab{c}})\BibitemShut {NoStop}%
\bibitem [{\citenamefont {{The Qiskit
  Team}}(2022{\natexlab{d}})}]{Qiskit_Gates}%
  \BibitemOpen
  \bibfield  {author} {\bibinfo {author} {\bibnamefont {{The Qiskit Team}}},\
  }\href@noop {} {\enquote {\bibinfo {title} {{Summary of Quantum
  Operations}},}\ }\bibinfo {howpublished}
  {\url{https://qiskit.org/documentation/tutorials/circuits/3_summary_of_quantum_operations.html}}
  (\bibinfo {year} {2022}{\natexlab{d}})\BibitemShut {NoStop}%
\bibitem [{\citenamefont {Tennie}\ \emph {et~al.}(2023)\citenamefont {Tennie},
  \citenamefont {Danin},\ and\ \citenamefont {Farrow}}]{Tennie_2023}%
  \BibitemOpen
  \bibfield  {author} {\bibinfo {author} {\bibfnamefont {F.}~\bibnamefont
  {Tennie}}, \bibinfo {author} {\bibfnamefont {D.}~\bibnamefont {Danin}}, \
  and\ \bibinfo {author} {\bibfnamefont {T.}~\bibnamefont {Farrow}},\
  }\href@noop {} {\  (\bibinfo {year} {2023})},\ \bibinfo {note}
  {{(Forthcoming)}}\BibitemShut {NoStop}%
\end{thebibliography}%

\end{document}